\documentclass[]{elsarticle}

\journal{}

\usepackage{amssymb}
\usepackage{booktabs}
\usepackage{subfigure}
\usepackage{multirow}
\usepackage{hyperref}

\usepackage[figuresright]{rotating}

\begin{document}

\begin{frontmatter}


\author[label1]{Simon Harding\corref{cor1}}
\ead{slh@evolutioninmaterio.com}
\ead[url]{www.evolutioninmaterio.com}

\author[label2]{Jan Koutn\'ik}
\ead{hkou@idsia.ch}
\ead[url]{http://people.idsia.ch/~koutnik/}

\author[label2]{Klaus Greff}
\ead{klaus@idsia.ch}
\ead[url]{http://people.idsia.ch/~greff/}

\author[label2]{J\'{u}rgen Schmidhuber}
\ead{juergen@idsia.ch}
\ead[url]{http://people.idsia.ch/~juergen/}

\author[label1]{Andy Adamatzky}
\ead{andrew.adamatzky@uwe.ac.uk}
\ead[url]{http://uncomp.uwe.ac.uk/adamatzky/}





\cortext[cor1]{I am corresponding author}
\address[label1]{Unconventional Computation Centre, University of the West of England, Bristol UK}
\address[label2]{IDSIA, USI\&SUPSI, Manno-Lugano, CH}


\title{Discovering Boolean Gates in Slime Mould}

\author{}

\address{}

\begin{abstract}
Slime mould of Physarum polycephalum is a large cell exhibiting rich spatial non-linear electrical characteristics. We exploit the electrical properties of the slime mould to implement logic gates using a flexible hardware platform designed for investigating the electrical properties of a substrate (\textit{Mecobo}). We apply arbitrary electrical signals to `configure' the slime mould, i.e. change shape of its body and, measure the slime mould's electrical response. We show that it is possible to find configurations that allow the Physarum to act as any 2-input Boolean gate. The occurrence frequency of the gates discovered in the slime was analysed and compared to complexity hierarchies of logical gates obtained in other unconventional materials. The search for gates was performed by both sweeping across configurations in the real material as well as training a neural network-based model and searching the gates therein using gradient descent.
\end{abstract}


\end{frontmatter}


\section*{Introduction}

Slime mould \emph{Physarum polycephalum} is a large single cell~\cite{stephenson1994myxomycetes} capable for distributed sensing, concurrent information processing, parallel computation and decentralized actuation~\cite{adamatzky2010physarum,adamatzkyAdvancesPhysarum}. The ease of culturing and experimenting with Physarum makes this slime mould  an ideal substrate for real-world implementations of unconventional sensing and computing devices~\cite{adamatzky2010physarum}.  A range of hybrid electronic devices were implemented as experimental working prototypes. They include Physarum self-routing and self-repairing wires \cite{adamatzky2013physarumwire}, electronic oscillators \cite{adamatzky2014slimeoscillator}, chemical sensor \cite{whiting2014towards}, tactical sensor \cite{adamatzky2013slime},  low pass filter \cite{whiting2015transfer},  colour sensor \cite{adamatzky2013towards}, memristor \cite{gale2013slime, tarabella2015hybrid}, robot controllers \cite{tsuda2006robot,galeAndroid}, opto-electronics logical gates \cite{mayne2015slimegates}, electrical oscillation frequency logical gates \cite{whiting2014slimefrequency}, FPGA co-processor \cite{mayne2015towardsFPGA}, Shottky diode \cite{cifarelli2014non}, transistor \cite{tarabella2015hybrid}.  There prototypes show that Physarum is amongst most
prospective candidates for future hybrid devices, where living substrates physically share space, interface with and co-function with conventional silicon circuits. 

So far there are four types of Boolean gates implemented in Physarum. Two of the prototypes employ propagation of Physarum: logical False is an absence of tube in a particular point of substrate and logical True is a presence of the tube \cite{tsuda2004robust, adamatzky2010slime}. One of the `morphological' gates employs chemotactic behaviour of Physarum \cite{tsuda2004robust}, another prototype is based on inertial propagation of the slime mould \cite{adamatzky2010slime}. Third prototype of Physarum logical gates is a hybrid device, where growing Physarum acts as a conductor and the Physarum's behaviour is controlled by light \cite{mayne2015slime}. Fourth prototype interprets frequencies of Physarum's electrical potential oscillations as Boolean values and input data as chemical stimuli \cite{whiting2014slime}. These logical circuits are slow and unreliable, and are difficult to reconfigure. Thus we aimed to develop an experimental setup to obtain fast, repeatable and reusable  slime mould circuits using the `computing in materio' paradigm. 

The Mecobo platform has been designed and built within an EU-funded research project called NASCENCE~\cite{BroersmaNAS2012}. The purpose of the hardware and software is to facilitate `evolution in materio' (EIM) --- a process by which the physical properties of a material are exploited to solve computational problems without requiring a detailed understanding of such properties \cite{Miller:2014}. 

EIM was inspired by the work of Adrian Thompson who investigated whether it was possible for unconstrained evolution to evolve working electronic circuits using a silicon chip called a Field Programmable Gate Array (FPGA). He evolved a digital circuit that could discriminate between 1\,kHz or 10\,kHz signal \cite{adrian_thompson:book:hardware_evolution}. When the evolved circuit was analysed, Thompson discovered that artificial evolution had exploited physical properties of the chip. Despite considerable analysis and investigation Thompson and Layzell were unable to pinpoint what exactly was going on in the evolved circuits \cite{thom:99}. Harding and Miller attempted to replicate these findings using a liquid crystal display \cite{harding04b}. They found that computer-controlled evolution could utilize the physical properties of liquid crystal --- by applying configuration signals, and measuring the response from the material --- to help solving a number of computational problems \cite{harding:ijuc:2007}. 

In the work by Harding and Miller, a flexible hardware platform known as an `Evolvable Motherboard' was developed. The NASCENCE project has developed a new version of this hardware based on low-cost FPGAs that more flexible than the original platform used with liquid crystal.

In this paper, we describe the Mecobo platform (\autoref{sec:mecobo}) and how it is interfaced to Physarum (\autoref{sec:inter}) to perform an exhaustive search over a sub-set of possible configurations. In  \autoref{sec:search} the results from this search are then searched to confirm whether Physarum is capable of acting as Boolean gates. We find that all possible 2-input gates can be configured this way. \autoref{sec:ann} further explores the complexity of the computation by modelling the material using a neural network model.

\section*{The Mecobo Platform}
\label{sec:mecobo}

Mecobo is designed to interface a host computer to a candidate computational substrate, in this instance the Physarum. 
The hardware can act as a signal generator, and route these signals to arbitrary `pins' --- which are typically electrodes connected to the candidate material. The hardware can also record the electrical response from the substrate. Again, this recording (or recordings) can be linked to any `pin'. Both applied signals (used for configuration) and measurements can be analog or digital.

\begin{figure}[!tbp]
\centering
\includegraphics[width=0.99\textwidth]{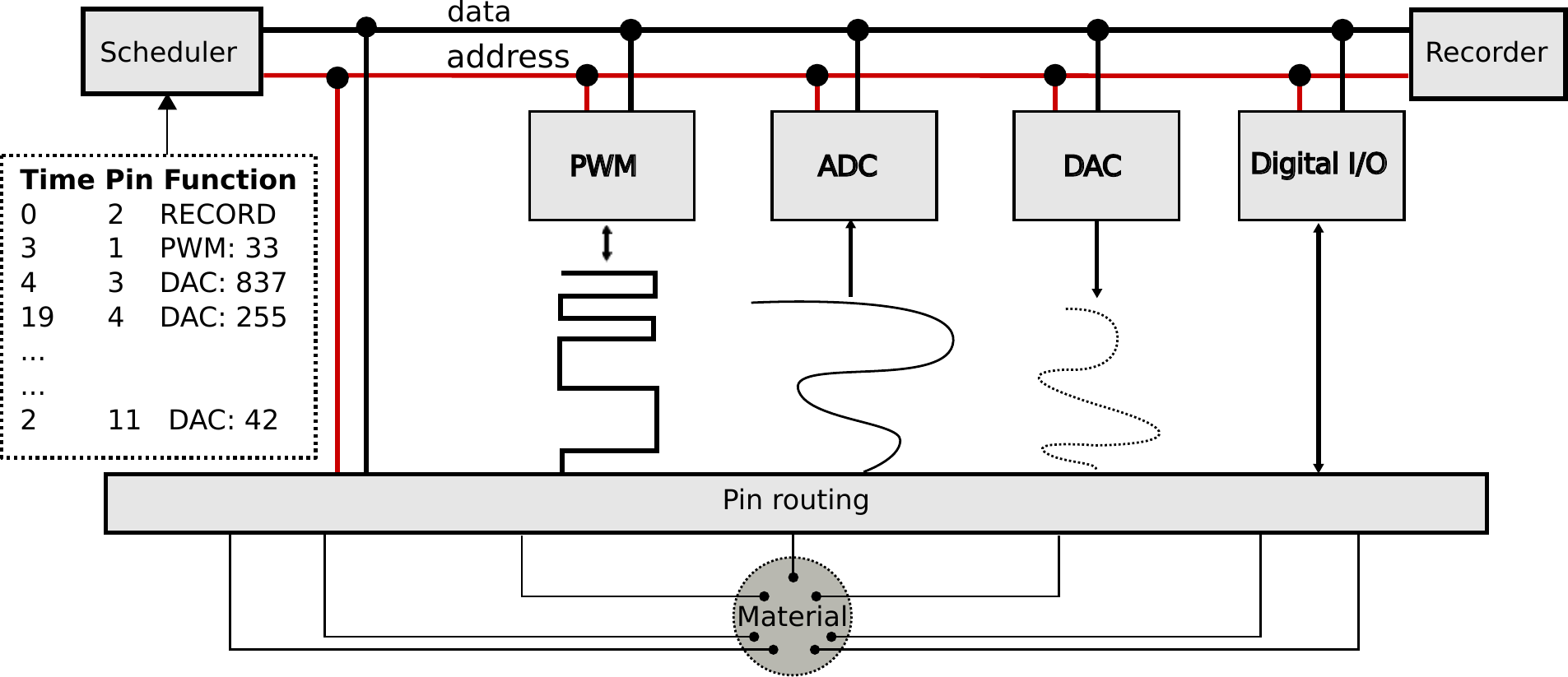} 
\caption{Overview of the low-level Mecobo hardware architecture.}
\label{fig:arch2}
\end{figure}

The core of the hardware is a microprocessor and an FPGA that run a `scheduler', illustrated in \autoref{fig:arch2}. A series of actions, such as ``output 1Hz square wave on pin 5, measure on pin 3'' are placed into a queue, and the queue executed. Complex series of actions can be scheduled onto the Mecobo. The Mecobo connects, via USB, to a host PC. The host PC runs software implementing the server side of the Mecobo Application Programming Interface (API). The API work flow mirrors the hardware's scheduler, and allows for applications to interface to the hardware without understanding all of the underlying technical details. Implemented using THRIFT, the API is also language and operating system agnostic, with the additional benefit that it can executed remotely over a network. The API software also includes functionality for data processing and logging of collected data for later analysis.

\begin{figure}[!tbp]
\centering
\includegraphics[width=0.99\textwidth]{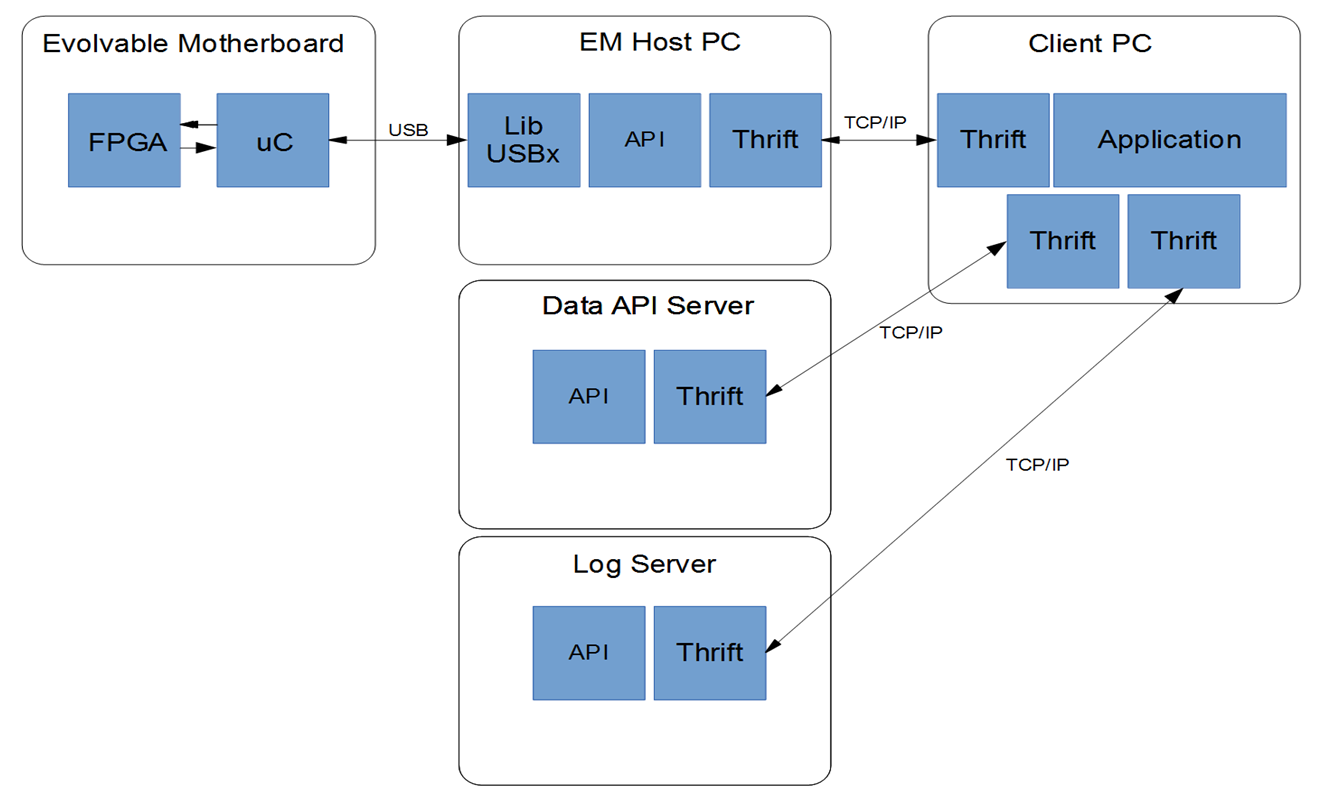} 
\caption{Overview of the Mecobo hardware software architecture.}
\label{fig:arch}
\end{figure}

\autoref{fig:arch} shows a simplified architectural overview of the Mecobo hardware and software. A full description, of both hardware and software can be found in \cite{oddrune2014}. The hardware, firmware and software for the Mecobo are open source and can be accessed at the project website: \url{http://www.nascence.eu/}.

\section*{Interfacing Physarum to Mecobo}
\label{sec:inter}

\begin{figure}
\begin{center}
\subfigure[]{\includegraphics[width=0.3\textwidth]{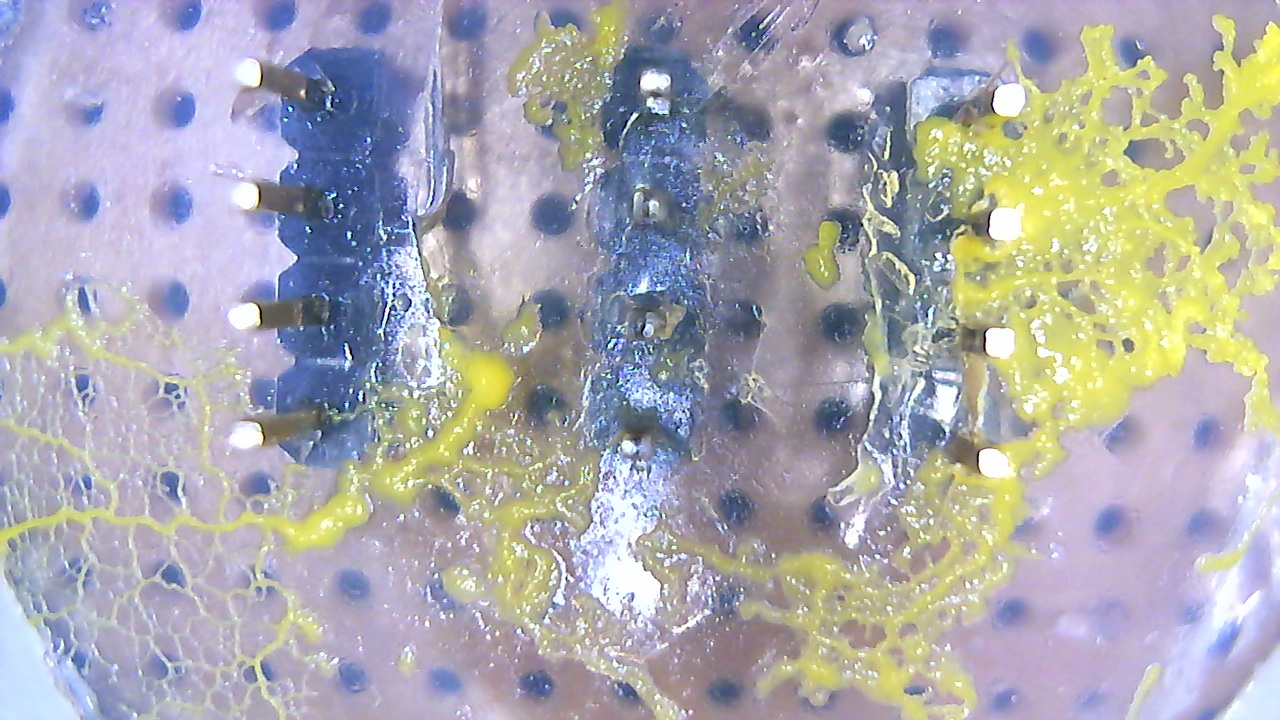}}
\hspace{2mm}
\subfigure[]{\includegraphics[width=0.3\textwidth]{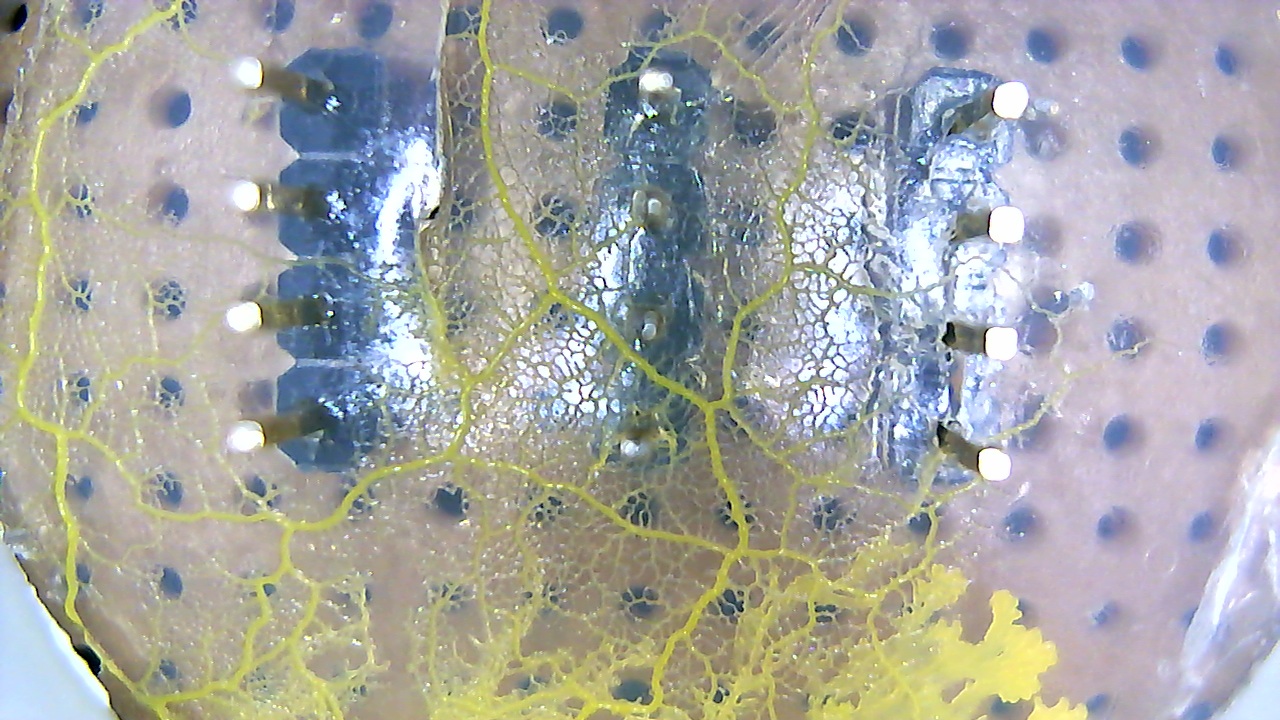}}
\hspace{2mm}
\subfigure[]{\includegraphics[width=0.3\textwidth]{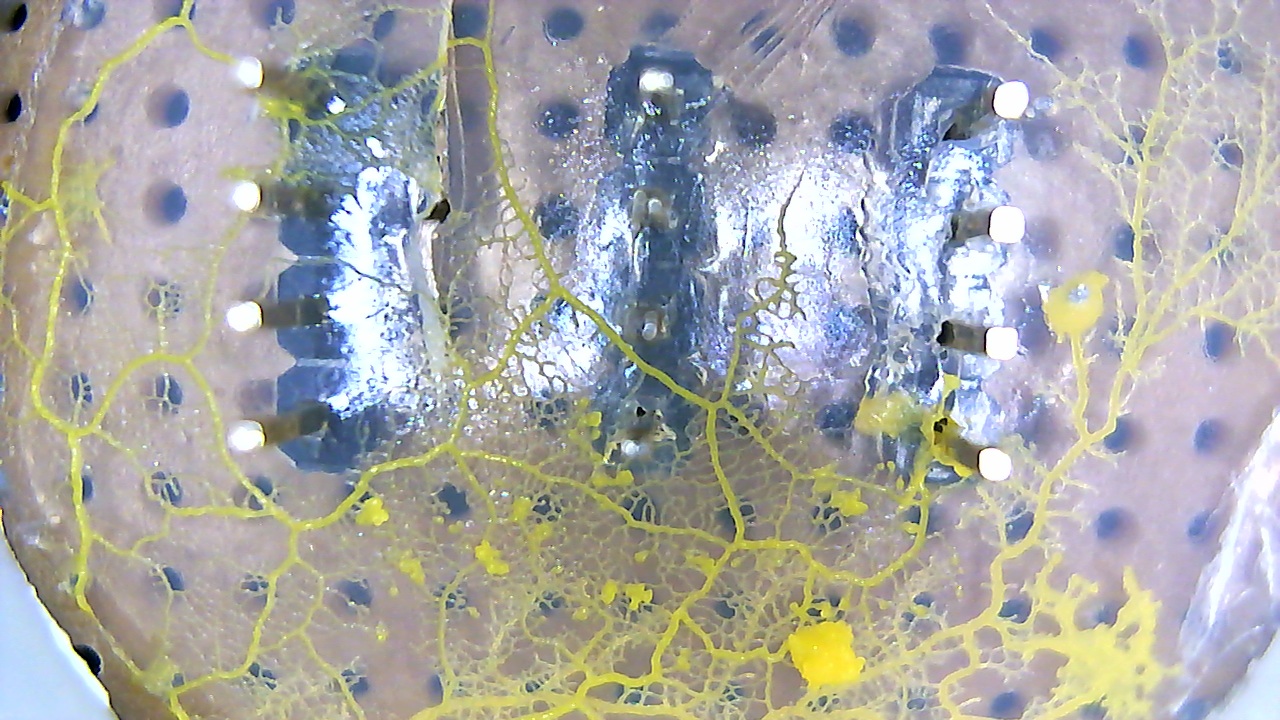}}\\
\subfigure[]{\includegraphics[width=0.3\textwidth]{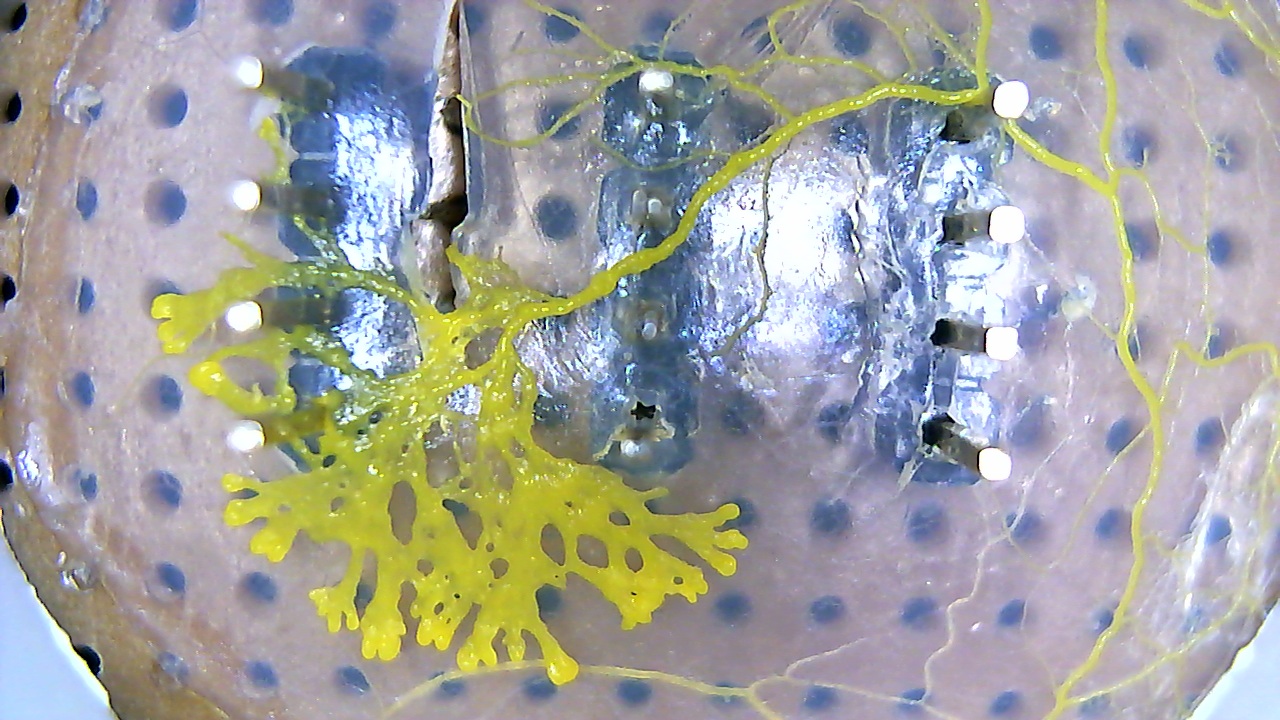}}
\hspace{2mm}
\subfigure[]{\includegraphics[width=0.3\textwidth]{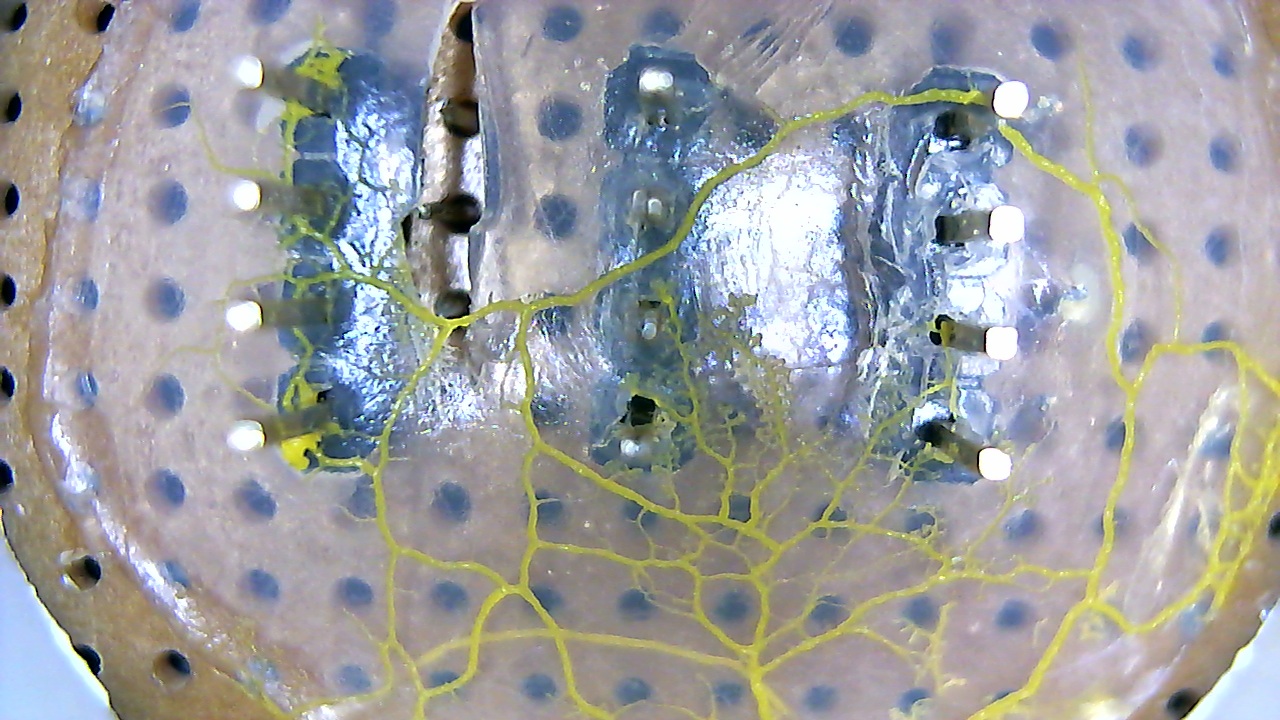}}
\hspace{2mm}
\subfigure[]{\includegraphics[width=0.3\textwidth]{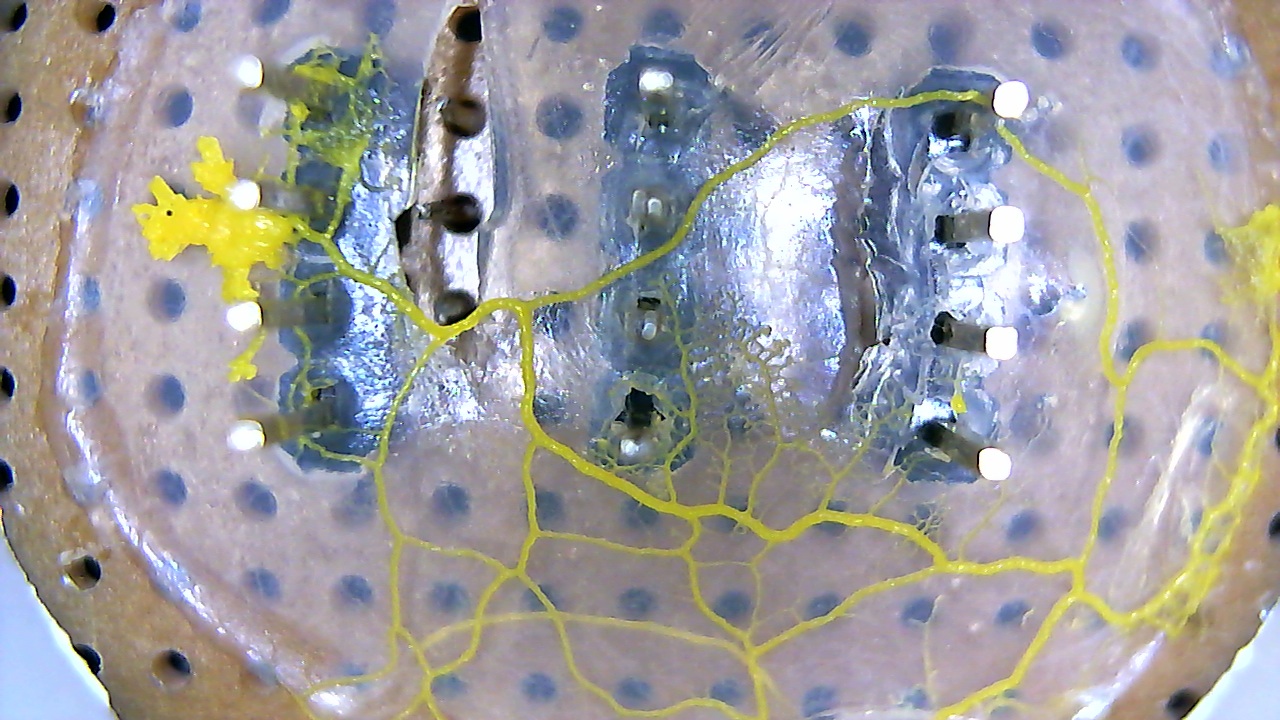}}
\caption{Time-lapse of Physarum growing on agar with electrodes. Images are approximately 6 hours apart.}
\label{fig:moldevo}
\end{center}
\end{figure}

\begin{figure}[!tbp]
\begin{center}
\subfigure[]{\includegraphics[width=0.4\textwidth]{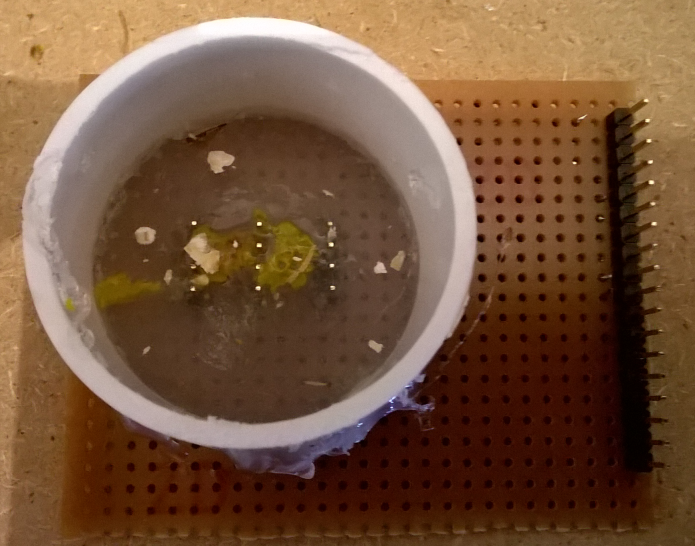}}\hspace{3mm}
\subfigure[]{\includegraphics[width=0.4\textwidth]{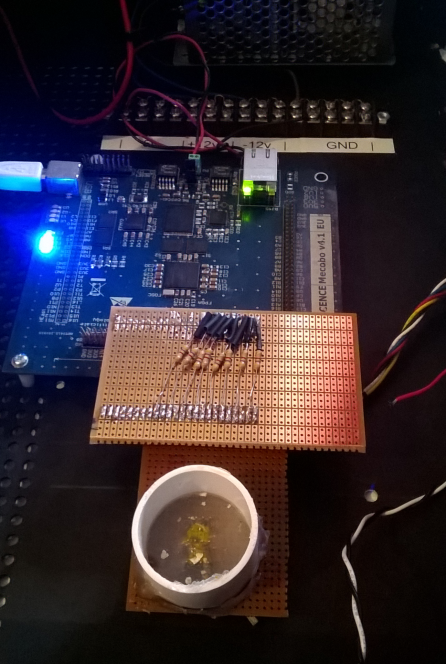}}
\caption{(a) Electrodes with agar and Physarum. The white spots are oat flakes, which are used to feed the Physarum. (b) Agar dish connected to Mecobo.}
\label{fig:moldcup}
\end{center}
\end{figure}

A crude containing dish with electrodes was constructed from matrix board and gold-plated header pins, as shown in \autoref{fig:moldcup}(a). Onto these pins, we pushed down a section of $2\%$ agar (with or without Physarum) so that the pins punctured the agar, and were visible above the surface. This allowed the Physarum to come into direct contact with the electrodes. For experiments where the substrate was intended to be only Physarum, a small amount of molten agar was painted over the contacts and some of the matrix board. This minimized the amount of agar, in order to reduce its influence. Small amounts of damp paper towel were placed around the container to maintain humidity.

A video available at \url{https://www.youtube.com/watch?v=rymwltzyK88} shows the Physarum growing and moving around the electrodes.

As shown in \autoref{fig:moldcup}(b) the electrodes were then connected to the digital outputs of the Mecobo. Each electrode was connected to the Mecobo interface using a 4.7\,k$\Omega$ resistor to limit the amount of applied current. The digital outputs of the Mecobo provide either $0$v (low) or $3.3$v (high).

\section{Data collection}

The possible configuration space for the Mecobo platform is vast, and it would be infeasible to attempt to apply all possible configurations to the Physarum. In the Nascence project, evolutionary algorithms are used as a practical method to search through the configuration space and find configurations that perform a particular task. Just as with the initial experiments in the Nascence project \cite{oddrune2014}, we start with an exhaustive search over a small configuration space. 

The exhaustive search procedure consisted of applying all possible binary combinations of various frequency pairs to 9 pins (which is a practical amount for time purposes).  One pin was used as an output from the material, with the other 8 pins acting as inputs. 

For each binary combination, each pair of frequencies was tried with one frequency representing a `low' and the other representing a `high' input.  The frequency pairs were combinations of square waves of either $(250$\,Hz, $500$\,Hz, $1$\,kHz, or $2.5$\,kHz$)$. 

The order of applied configurations was shuffled to help prevent similar configurations being applied sequentially.

The Mecobo measured the digital response from the Physarum and/or agar for 32ms. The digital threshold is $0.75$v for high, with voltages below this being classed as low. The sampling frequency was twice the highest input frequency applied.

In total, $49\,152$ states are applied, in a process that takes approximately $2$ hours to complete. The majority of the time is spent configuring the Mecobo and transferring sampled data back to the host PC.

\section*{Exhaustive search for Boolean gates}
\label{sec:search}
\subsection{Method}

Boolean logic gates are the building block for many circuits, sensors and computational devices. Using an exhaustive search, we were able to data mine within the collected data to look for configurations that acted as Boolean gates.  

The data was searched for all two-input Boolean functions. Two pins were selected as the inputs to the gates ($A$ and $B$), and one pin was selected as the gate output. The output response of the material (i.e. \emph{true} or \emph{false}) was determined from the output frequency of the material. A Fast Fourier Transform (FFT) was performed, and the frequency of the highest magnitude response was determined. The output was \emph{true} if this frequency was nearer to the \emph{true} input frequency, and \emph{false} if nearer in value to the \emph{false} input frequency. The remaining pins were then treated as configuration pins for the Physarum/agar. Each possible combination of configuration, input pair and output were then compared to see what logical operation it performed.

If for a given configuration (using the same frequency pairs), the output for all combinations of inputs $A$ and $B$ matched the expected behaviour of a gate, then it was judged that the Physarum and/or agar could implement that gate. As the configurations are temporally spaced, and each configuration applied multiple times (as $A$ and $B$ could be swapped over), a discovered gate would likely be `stable' over the time taken to run the experiment. As the Physarum grows and moves, or as the agar dehydrates and shrinks, the physical substrate will vary. Therefore, we would not expect the system to be stable over long periods of time. Repeated measurements that are temporally spaced also reduces the possibility of measurement noise strongly influencing the results.

\subsection{Results}

\begin{table}[!tbp]
\begin{center}
\caption{Number of gates mined from the frequency responses of the Physarum. The {\it Agar} 
column contains a number of such configurations (input pins having values \emph{true}, T, or \emph{false}, F) found in the bare agar substrate and the  {\it Physarum} column contains number of the configuration found 
in the dish containing the Physarum. We can see that the Physarum actually performs the logic functions whereas the sole agar is not capable of that.
}
\label{tab:mold:gates}
\addtolength{\tabcolsep}{-3pt}
\begin{tabular}{ccccccccc}
\toprule
{\bf Cfg.}& \multicolumn{4}{c}{Inputs $xy$} & {\bf Agar} & {\bf Physarum} & {\bf Physarum} &{\bf Gate} \\
   & FF & FT & TF & TT &  &{\bf \& Agar}  & &  \\
\midrule
1 &  F  &  F  &  F  &  F  &  76\,104  &  121\,144  & 15864 &  Constant False\\
2 &  T  &  F  &  F  &  F  &    & 86 & 8 &  $x$ NOR $y$ \\
3 &  F  &  T  &  F  &  F  &    & 201 & 13 &  NOT $x$ AND $y$\\
4 &  T  &  T  &  F  &  F  &    & 60 & 2 &  NOT $x$\\
5 &  F  &  F  &  T  &  F  &   & 201 & 13 &  $x$ AND NOT $y$\\
6 &  T  &  F  &  T  &  F  &   & 60 & 2 &  NOT $y$\\
7 &  F  &  T  &  T  &  F  &   & 66 & 4 &  $x$ XOR $y$ \\
8 &  T  &  T  &  T  &  F  &   & 112 & 10 &  $x$ NAND $y$ \\
9 &  F  &  F  &  F  &  T  &  43\,546  &  13\,268  & 564 &  $x$ AND $y$ \\
10 &  T  &  F  &  F  &  T  &    & 52 & 0 &  $x$ XNOR $y$ \\
11 &  F  &  T  &  F  &  T  &  15\,249  &   18\,259 & 3\,707 &  $y$\\
12 &  T  &  T  &  F  &  T  &    & 260 & 9 &  NOT $x$ AND NOT $y$ OR $y$\\
13 &  F  &  F  &  T  &  T  &  15\,249  &   18\,259 & 3\,707 &  $x$\\
14 &  T  &  F  &  T  &  T  &   & 260 & 9 &  $x$ OR NOT $y$\\
15 &  F  &  T  &  T  &  T  &  43\,564  &  13\,128  & 536 &  $x$ OR $y$ \\
16 &  T  &  T  &  T  &  T  &  74\,996  &  113\,266  & 13\,448 &  Constant True\\
\bottomrule
\end{tabular}
\addtolength{\tabcolsep}{3pt}
\end{center}
\end{table}

As detailed in Table \ref{tab:mold:gates}, agar on its own was unable to produce the universal gates NAND or NOR. It was also unable to produce the non-linear gates XOR or NXOR. Substrates that involve Physarum can be seen to produce more types of gates. It is particularly interesting to note that far fewer AND or OR gates were found when using Physarum+Agar, compared to only agar.
  
Using Physarum, both with the agar and with minimal agar, the search was able to find many types of logic gates. 

When the amount of agar was minimised, the fewest gates were found. This result was expected as the Physarum did not appear to connect between many of the electrodes, and therefore would only be able to participate in a smaller number of the configurations.

\subsubsection{Detailed Analysis XOR}

Considering only the XOR solutions that occur in the Physarum+Agar substrate, we can investigate the collected data in some detail.
Table \autoref{tab:mold:xor} shows the frequency of times that particular pins are used as inputs (the symmetry is a result of the fact that in this instance A and B can be swapped and still produce a valid gate). 

\begin{table}[!tbp]
\begin{center}
\caption{Number of XOR gates found for given Input pin configurations.
}
\label{tab:mold:xor}
\begin{tabular}{cccccccccc}
\toprule
  &  & \multicolumn{8}{c}{\bf Input Pin A}  \\
  &  & {\bf 0} & {\bf 2}& {\bf 3} & {\bf 4} & {\bf 5} & {\bf 6} & {\bf 7} & {\bf 8} \\
\midrule
\multirow{8}{*}{\bf \rotatebox[origin=c]{90}{Input Pin B}} & {\bf 0} &  & 24 & 24 & 16 & 40 & 24 &  & 8 \\
  & {\bf 2} & 24 &  &  & 16 & 8 &  &  &  \\
  & {\bf 3} & 24 &  &  &  & 8 &  & 16 &  \\
  & {\bf 4} & 16 & 16 &  &  & 24 & 8 & 16 & 8 \\
  & {\bf 5} & 40 & 8 & 8 & 24 &  & 8 &  &  \\
  & {\bf 6} & 24 &  &  & 14 & 8 &  & 8 &  \\
  & {\bf 7} &  &  & 16 & 16 &  & 8 &  &  \\
  & {\bf 8} & 8 &  &  & 8 &  &  &  &  \\
\bottomrule
\end{tabular}
\end{center}
\end{table}

\autoref{tab:mold:xor} shows that some electrode pins are used much more frequently than others as inputs, and that only two electrodes ever get used successfully as outputs. This strongly suggests that the physical structure of the Physarum is important, and that a uniform mass would not be effective.

\begin{figure}[!tbp]
\begin{center}
\includegraphics[width=0.7\textwidth]{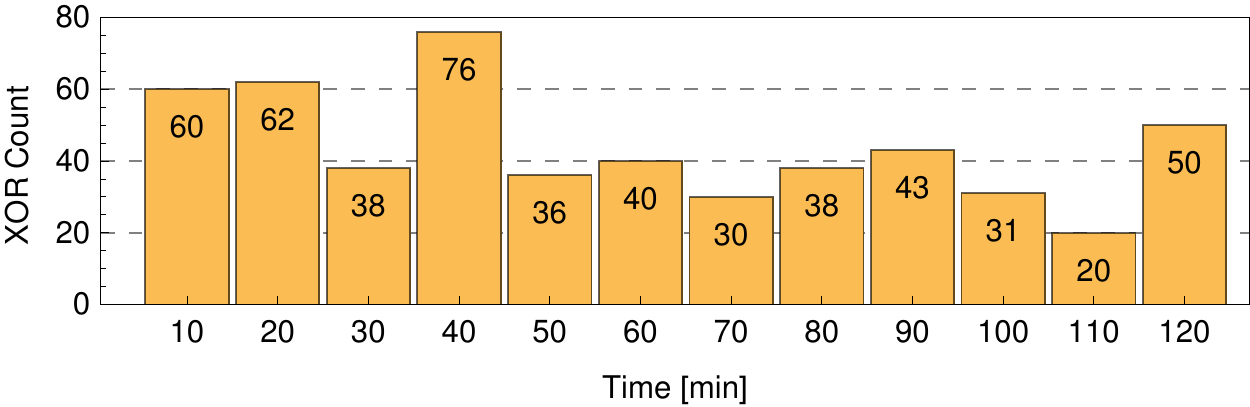}
\caption{XOR density. The chart shows how many XOR functions were found in the Physarum material over the course of 120 minute long experiment. We can see that most of the gates was found in the beginning.}
\label{fig:mold:hist}
\end{center}
\end{figure}

The experiment ran for $7\,322$ seconds ($122$ minutes). During this time we observed the Physarum did move around on the agar.  We generated a histogram that shows the number of times an applied configuration was used as part of XOR against time. The chart in \autoref{fig:mold:hist} indicates that most of the results were found earlier on in the experiment. In addition to the movement of the Physarum, this may be caused by the agar drying out and becoming less conductive. It therefore seems likely that there might be more possible gate configurations, but that the search was too slow to discover them before the characteristics of the Physarum+agar changed.

\section*{Mould Modelling using Neural Networks}
\label{sec:ann}

This section uses neural network (NN;~\cite{haykin:1998:NNC:521706}) models, trained by gradient descent \cite{Linnainmaa:1970,dreyfus:1973,Werbos:74} to approximate the mapping from input voltages to the
output voltage by training them on many randomly chosen examples measured during the exhaustive search.
By solving that task a NN becomes a differentiable model of the---potentially complex---structures inside the material. Having the NN model, one can assess the complexity of the material based on trainability of the NN model.

A NN consist of a sequence of layers, where each layer computes an affine projection of its inputs followed by the application of a point-wise non-linear function $\psi$: 
\[
\mathbf{h}=\psi(\mathbf{Wx+b})
\]
Where $\theta = \{\mathbf{W, b}\}$ are the parameters of the layer. 
By stacking these layers one can build non-linear functions of varying expressiveness that are differentiable. In theory they can approximate any function to arbitrary precision given enough hidden units.

\subsubsection{Data Preprocessing}

The random search collected data was preprocessed in order to form a supervised training set. 
The input frequencies were ordered and translated to integer numbers in range from 
$1$ to $5$ forming an input vector $\mathbf{x}\in\{1,2,3,4,5\}^{9}$, where the input frequency was present. The value of $0$ was used in the case of a grounded input pin or when the pin was used as the output pin. The target output vector had $9$ entries as well ($\mathbf{t}\in[0,1]^{9}$) but only the one that corresponds to the output pin was used at a time.

A {\it middle} section of the output buffer (1/4 to 3/4 section of the signal length) was preprocessed and transformed to a desired output target in one following three ways in the three experiments:

\begin{itemize}
\item {\bf Ratio}: The proportion of ones in the signal.
\item {\bf Peak Frequency}: The peak of the frequency spectrum normalised such that $0.0$ corresponds to a constant signal and $1.0$ corresponds
to the sequence $010101\dots$. The peak was obtained by computing the Fourier spectrum of the middle section of the signal, removing the DC component (first frequency component), computing the absolute values of the complex spectrum and using the first half (the spectrum is symmetric) to find the position of the maximum peak index.
\item {\bf Compressibility}: Relative length of the output buffer compressed using
LZW to the maximum length of the compressed buffer. High values point
to irregular outputs.
\end{itemize}


The networks have been trained to predict
the output vector given the inputs, where we only trained for the
one active output while ignoring the other 8 predictions of the network.
Note that this essentially corresponds to training 9 different neural
networks with one output each, that share the weights of all but their
last layer. 

The network weights were trained using Stochastic Gradient Descend (SGD) with a minibatch-size of 100 to minimise the Mean Squared Error
on the active output. We kept aside 10\% of the data as a validation
set. Training was stopped after 100 epochs (full cycles through the
training data) or once the error on the validation set didn't decrease
for 5 consecutive epochs. 

Neural networks have certain hyperparameters like the learning rate
for SGD and the network-architecture that need to be set. To optimize
these choices we performed a big random search of 1650 runs sampling
the hyperparameters as follows:
\begin{itemize}
\item learning rate $\eta$ log-uniform from $10^{-3}$ to $10^{-1}$
\item number of hidden layers uniform from \{1, 2, 3, 4, 5, 6, 7, 8\}
\item number of hidden units in each layer from the set \{50, 100, 200,
500\}
\item The activation function of all hidden units from the set \{tanh, ReLU\footnote{ReLU: Rectified Linear Unit},
logistic sigmoid\}
\end{itemize}
The best networks for the three tasks can be found in \autoref{tab:bestnets}.

\begin{table}
\begin{centering}
\protect\caption{Best hyperparameter settings for each of the three tasks along with
the resulting Mean Squared Error (MSE).}
\label{tab:bestnets}
\addtolength{\tabcolsep}{-3pt}
\begin{tabular}{cccccc}
\toprule 
{\bf Task} & {\bf \# Layers} & {\bf \# Units} & {\bf Act. Fn.} & {\bf Learning Rate} & {\bf MSE}\tabularnewline
\midrule 
{\bf Ratio} & 8 & 200 & ReLU & 0.055 & $6.058\times 10^{-4}$\tabularnewline
{\bf Frequency} & 5 & 200 & ReLU & 0.074 & $2.68\times 10^{-2}$\tabularnewline
{\bf Compressibility} & 5 & 200 & ReLU & 0.100 & $7.09\times 10^{-4}$\tabularnewline
\bottomrule 
\end{tabular}
\addtolength{\tabcolsep}{-3pt}
\end{centering}
\end{table}


\begin{figure}[!tbp]
\begin{centering}
\includegraphics[width=\textwidth]{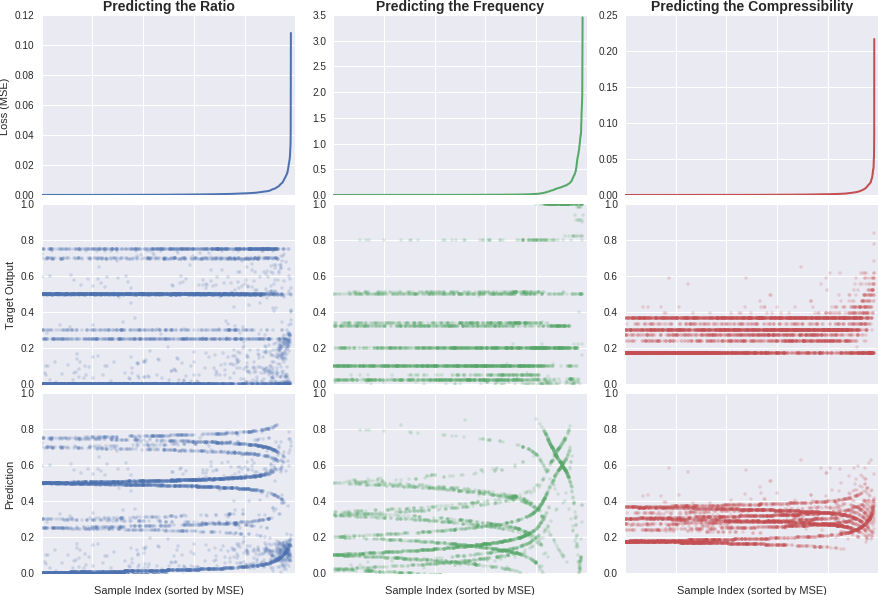}
\par\end{centering}
\protect\caption{Loss, true output, and prediction for all validation samples on each
of the three tasks. The examples are sorted by their associated training
loss. }
\label{fig:val_errors}
\end{figure}

\subsubsection{Modelling Results}

On all three tasks the networks prediction error on the validation
set decreased significantly during training. \autoref{fig:val_errors}
depicts the loss, the target and the prediction of the network for
each example of the validation for all three tasks sorted by their
loss. It can be seen that in all three cases the targets (\emph{true} outputs)
have a characteristic structure consisting of horizontal lines. This
is due to the fact that certain target values are very likely. For
each such value there are easy cases that the networks predict with
high accuracy (the left side of the bar), and there seem to be difficult
cases (right part of the bars) for which the predictions bifurcate
(see prediction plots at the bottom of \autoref{fig:val_errors}). This
essentially means that the output distribution has sharp peaks which
get smoothed out a bit by the networks. Overall, the networks predictive
performance is very good (apart from the high values in the frequencies
task which all fall into the high-loss region). The worst performance
is on the frequencies task, while the best performance is achieved
on the first task (predicting the ratios). 

\subsubsection{Searching for Logic Gates in the NN Model}

The best network for the first task (predicting
the ratio) was searched for the following logic functions: 
AND, OR, NAND,
NOR, XOR, and XNOR. We encode the values for \emph{True} and \emph{False} for the
inputs as 4 and 1 respectively (corresponding to high and low input
frequency). For the output values we define 0 as \emph{False} and 0.5 as
\emph{True}, because these two values are easy to distinguish and are the
most common output values. 


First, a set of examples with different combinations of
input values but which share the same (random) values for the configuration
pins along with the desired output values.  

Gradient descent is then used to minimise the MSE by adjusting the
values for the configuration pins, but we keep their values the same
for all examples. So formally, given our neural network model $\hat{f}$
of the nano-material, we define an error over our $N$ input/output
pairs $(I_{1}^{(i)},I_{2}^{(i)},O^{(i)})$: 
\[
E=\sum_{i=1}^{N}\frac{1}{2}(\hat{f}(I_{1}^{(i)},I_{2}^{(i)},\theta)-O^{(i)})^{2}
\]
The gradient is then calculated using backpropagation of the error: 

\[
\frac{\partial E}{\partial\theta}=(\hat{f}(I_{1}^{(i)},I_{2}^{(i)},\theta)-O^{(i)})\frac{\partial\hat{f}}{\partial\theta}
\]
where $\theta=\{c_{1},\dots,c_{6}\}$. 

Unfortunately, at the end of this gradient descent process the configuration
pin values won't be in the allowed set \{1, 2, 3, 4\} anymore. To
make them valid inputs we round and clip them. This will most likely
increase the error but hopefully still remain in a region where the
network computes the desired function. It is unlikely that this procedure
will always work, so in addition to this local search we also perform
a global search. 

One problem with the method described above is that it only performs
a local search, which means that the solution that it converges to,
might correspond to a bad local minimum. Furthermore the discretization
we need to perform at the end of the local search might lead to a
configuration which doesn't approximate the desired function very
well. 

\begin{figure}[!tbp]
\begin{centering}
\includegraphics[width=\textwidth]{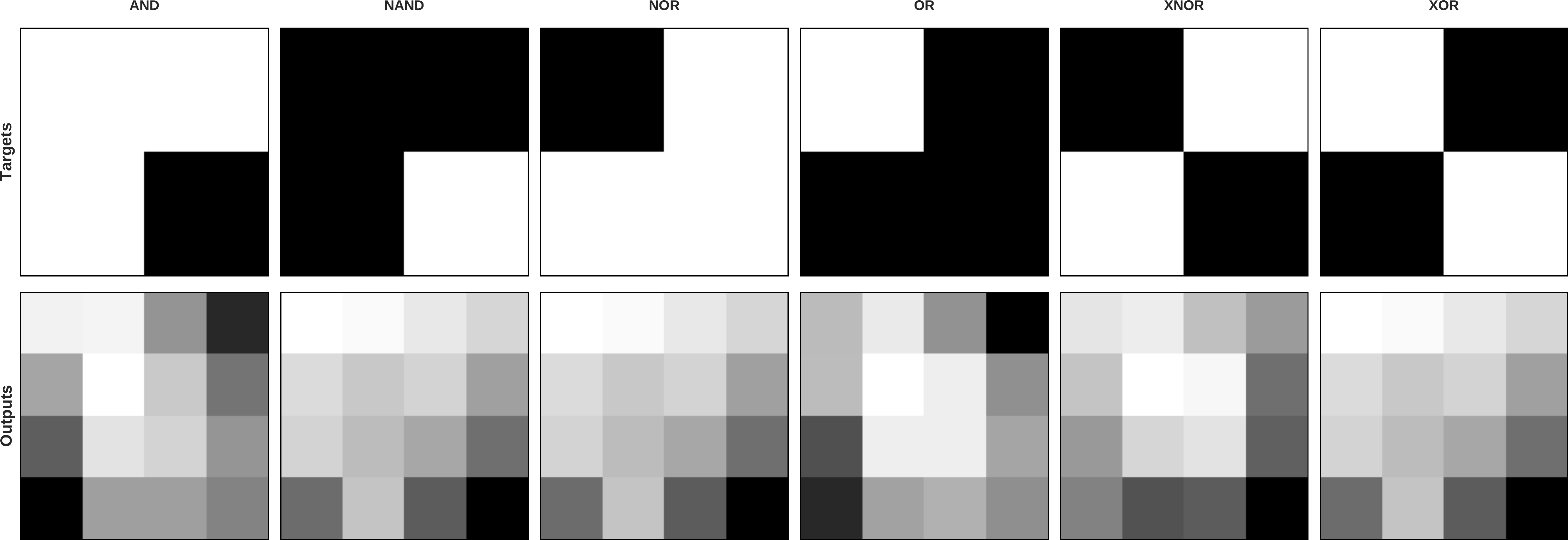}
\par\end{centering}

\protect\caption{The functions found by our search procedure all with pins 1 and 2 for
inputs and pin 4 as output. The results for AND and NOR are a good
fit, while the search for OR and XOR clearly failed to produce usable
functions. }
\label{fig:slimeresults}
\end{figure}

To mitigate these problems we first sample $1\,000$ random starting points
(settings of the configuration leads) and perform just 10 iterations
of our local search on them. Only the starting point that lead to
the lowest error is then optimised further for 500 epochs to obtain
the final solution. In this way we reduce the risk of getting stuck
in a poor local minimum. There is another free parameter which we
haven't optimised yet, which is the assignments of input and output
pins. We therefore repeat the above procedure for all 224 possible
allocations of input and output pins. The found gates and their 
responses are summarised in \autoref{fig:slimeresults}.

\section{Discussion}

We do not know what is exact physical mechanism of the gate operations implemented by the slime mould. We can speculate that the following factors might contribute to the particular responses observed:
\begin{itemize}
\item The slime mould is conductive. Every single protoplasmic tube can be considered as a wire \cite{adamatzky2013physarumwire}.
\item The resistance of the slime mould wire depends on biochemical and physiological state of slime, thickness of tubes, and the Physarum's reaction to environmental conditions. The resistance also oscillates, due to contractile activity and shuffling of cytoplasm in the protoplasmic tubes \cite{adamatzky2014slimeoscillator}
\item Physarum exhibits memristive properties when tested in current-voltage cycles \cite{gale2013slime}.  Thus some pathways in Physarum could become non-conductive under AC stimulation. 
\item Morphology of Physarum can be modified by application of AC~\cite{tsuda2012routing}. That is the slime mould can physically move between electrodes in response to stimulation by input electrical signals.
\item In certain conditions Physarum allows the conduction of AC frequencies with attenuation profile similar to a low pass filter \cite{whiting2015transfer,whiting2015practical}. 
\end{itemize}

\noindent
The frequency of gates' discovery in Physarum varies substantially between the gates, with XOR and XNOR being the most rarely observable gates. Is it typical for all non-linear systems implementing gates or just Physarum? Let gates $g_1$ and $g_2$ are discovered with occurrence frequencies  $f(g_1)$ and $f(g_2)$, we say gate $g_1$ is easy to develop or evolve than gate $g_2$: $g_1 \rhd g_2$ if  $f(g_1) > f(g_2)$. The hierarchies of gates obtained using evolutionary techniques in liquid crystals  \cite{harding2005evolution}, light-sensitive modification of  Belousov-Zhabotinsky system  \cite{toth2008dynamic} are compared with Physarum generated gates and morphological complexity of configurations of one-dimensional cellular automata governed by the gates \cite{adamatzky2009complex}: 
\begin{itemize}
\item Gates in liquid crystals \cite{harding2005evolution}: 
\{OR, NOR\} $\rhd$ AND $\rhd$ NOT $\rhd$ NAND $\rhd$ XOR
\item Gates in Belousov-Zhabotinsy medium \cite{toth2008dynamic}:
AND $\rhd$ NAND $\rhd$ XOR
\item Gates in cellular automata \cite{adamatzky2009complex}: 
OR $\rhd$ NOR $\rhd$ AND $\rhd$ NAND $\rhd$ XOR
\item Gates in Physarum: 
AND $\rhd$ OR $\rhd$ NAND $\rhd$ NOR $\rhd$ XOR $\rhd$ XNOR
\end{itemize}
We see that in all systems quoted the gate XOR is the most difficult to find, develop or evolve. Why is it so? A search for an answer could be one of the topics for further studies. 

The search of gate configurations in the differentiable material model renders to be an efficient method
of finding the correct configurations, that generates robust solutions. It remains to be confirmed whether these configurations work at the real mold again and close the loop between the modelling and the real material.


\section*{Acknowledgements}

The research leading to these results has received funding from the EC  FP7 under grant agreements 317662 (NASCENCE project) and 316366 (PHYCHIP project).

The authors would like to acknowledge the assistance of Odd Rune Lykkeb{\o} for his technical assistance with Mecobo. Simon and Andy prepared the mould and performed the exhaustive search experiments on the hardware platform, Jan, Klaus and J\"{u}rgen contributed with the mould neural network modelling. 


\section*{References}
%
%
%
\bibliographystyle{elsarticle-num}
\bibliography{bib}

\end{document}